\documentclass[manuscript]{aastex}

\shorttitle{Near-infrared background}
\shortauthors{Matsumoto et al.}

\begin{document}

\title{Reanalysis of the near-infrared extragalactic background light based on the IRTS observations}

\author{T. Matsumoto\altaffilmark{1}}
\affil{Institute of Astronomy and Astrophysics, Academia Sinica, Taipei 10617, Taiwan}

\author{M. G. Kim}
\affil{Seoul National University, Seoul 151-742, Korea}

\author{J. Pyo}
\affil{Korea Astronomy and Space Science Institute, Daejeon 305-348, Korea}

\and

\author{K. Tsumura}
\affil{Frontier Research Institute for Interdisciplinary Science,
Tohoku University, Sendai, Miyagi 980-8578, Japan}

\altaffiltext{1}{Institute of Space and Astronautical Science, Japan 
Aerospace Exploration Agency, Kanagawa 252-5210, Japan}

\begin{abstract}

We reanalyze data of near-infrared background taken by Infrared Telescope in Space (IRTS) based on up-to-date
observational results of zodiacal light, integrated star light and diffuse Galactic light.
We confirm the existence of residual isotropic emission, which is slightly lower but almost the same as 
previously reported.
At wavelengths longer than 2 $\mu$m, the result is fairly consistent with the recent observation with AKARI.
We also perform the same analysis using a different zodiacal light model by Wright and
detected residual isotropic emission that is slightly lower than that based on the 
original Kelsall model. Both models show the residual isotropic emission that is significantly brighter than the  integrated light of galaxies. 

\end{abstract}


\section{Introduction}

The extragalactic background light (EBL) has been observed over a wide range of wavelengths to
examine the energy density of the universe.  In particular, the near-infrared EBL has been thought to provide 
an important clue to our understanding of the early universe and the evolution of galaxies.
The COsmic Background Explorer (COBE)  \citep{Cambrecy01, Gorjian00,  Wright00, Levenson07}
and the InfraRed Telescope in Space (IRTS) (\citet{Matsumoto05}, hereafter referred to as paper I) discovered that a
significant fraction of the near-infrared isotropic emission cannot be explained with known foreground emission.
Recent AKARI observations \citep{Tsumura13d} also 
show a consistent result with COBE and IRTS at wavelengths longer than 2 $\mu$m.
This excess background emission in the near-infrared sky is particularly  interesting in light of the
recent discovery of large excess fluctuation of the near-infrared sky \citep{Kashlinsky05, 
Kashlinsky07a, Kashlinsky07b, Matsumoto11, Zemcov14}.

  The results of paper I attracted wide interest due to the high accuracy enabled 
by the unique low resolution spectroscopy of IRTS, and its point source 
detection limit ( $\sim$ 11 mag) which was much deeper than COBE.
However, as \citet{Mattila06} pointed out, paper I did not take the contribution of the diffuse 
 Galactic light (DGL) into account.
Uncertainty of the zodiacal light (ZL) model has been also raised, since ZL is the dominant 
foreground emission with a spectrum  similar to the residual isotropic emission \citep{Dwek05}.

In response to these concerns, we decided to reanalyze the IRTS data using 
up-to-date observations of ZL, integrated star light (ISL), and DGL. 
While in paper I, we adopted the  \citet{Kelsall98} ZL model, in this work, we perform the same analysis  
using a different ZL model (the so called Wright model,  \citet{Wright98}),
 and examine the difference between two models.

The overall outline of the paper is as follows. In section 1, we briefly present the IRTS observation and the acquisition of
raw data in section 2. In section 3, we estimate the contribution
of foreground emission, ZL, ISL and DGL 
based on the latest observations. In section 4, we search for the residual isotropic emission for two ZL models
based on the correlation with the sky brightness after subtracting the ISL and DGL. 
Finally, in section 5, we discuss the astrophysical implications of the detected excess brightness.

\section{IRTS observations}

We now give a  brief
description of the IRTS mission and the data acquisition process. Details of IRTS mission 
can be found in paper I.

IRTS was one of the mission experiments on the small space platform, Space Flyer Unit (SFU), that 
was launched on March 18, 1995. On a low-inclination near-earth 
orbit, IRTS continuously surveyed the sky avoiding both the sun and the earth. IRTS observations 
lasted for about 30 days, during which 7\% of the sky was surveyed \citep{Murakami96}. 

The Near InfraRed Spectrometer (NIRS) is one of the focal plane instruments of IRTS, and was optimized to 
obtain spectra of the diffuse background \citep{Noda94}. Details of the  flight performance of NIRS can 
be found in \citet{Noda96}. NIRS covered a wavelength 
range from 1.4 $\mu$m to 4.0 $\mu$ms with a spectral resolution of 0.13 $\mu$m,
providing 24 independent wavelength bands. 
The beam size was 8 arcmin square and the detection limit for
point sources was $\sim$11 mag, and both are considerably better than those of COBE.
The limiting magnitudes at wavelengths shorter and longer than 2.5 $\mu$m are 
dominated by the sky fluctuation and readout noise, respectively (\citet{Noda96}, paper I).

The NIRS detectors were of the charge integrating type and the ramp curves with one cycle of 65.54 sec were sent to 
the ground for all of the 24 wavelength bands. We retrieved the signal of the sky using 5 sec integration times
(i.e. the signal difference in 5 sec in ramp curve), during which no distinguishable stars 
and no cosmic ray hits were detected in any band. 
We obtained signals of the sky brightness after subtracting the dark current 
when the cold shutter was closed.
During each 5 sec integration, the telescope axis moved about 20 arcmin along a great circle, 
resulting in a trapezoidal beam pattern, 8 arcmin $\times$ 20 arcmin in area.  
To make the contribution of stars and Galactic emission less effective, high 
 Galactic latitude ($ b>40^{\circ}$) data were extracted from 
the full data set for the background radiation analysis.
The highest Galactic latitude was $58^{\circ}$, while the ecliptic latitude ranged from 
$12^{\circ}$ to $71^{\circ}$ in the selected sky.
Complete spectra of the sky were secured at 1010 fields. Sky coverage of 1010 fields
is  $\sim$ 60\% of the surveyed area.

\section{Foreground emission}

\subsection{Zodiacal light (ZL) }
The zodiacal light (ZL) is the emission component of the solar system which consists of
scattered sunlight and thermal emission by interplanetary dust.
We adopted the model by \citet{Kelsall98} in paper I, which is a physical model constructed
using the seasonal variation of the ZL observed with the Diffuse InfraRed Background Explorer (DIRBE) on COBE. 
At the same time,  \citet{Wright98} proposed a different physical model based on the so 
called "strong zodi principle" assuming no residual 
emission at 25 $\mu$m towards the ecliptic pole. In this paper, we use both models and examine the 
difference of residual isotropic emission. As for the Wright model, we used the model revised by \citet{Gorjian00}.

We retrieve the brightness of the ZL in the DIRBE bands corresponding to the 1010 IRTS fields at the epoch of the IRTS observations 
for both models, and construct the ZL model brightness for the IRTS bands. 
We obtain the model spectrum  for the scattering and thermal emission part, separately.  
For the scattering part, we simply normalize the spectral shape of the sun (ASTM G173-03 Reference System)
\footnote{Available in electronic form at  \\
http://rredc.nrel.gov/solar/spectra/am1.5/astmg173/astmg173.html} 
at the K band model brightness.
For the thermal part, we extrapolate the M band model brightness to the shorter IRTS 
wavelengths,  assuming a 300K blackbody in accordance with recent AKARI observations \citep{Tsumura13b}.
The ZL model spectrum is obtained by summing the scattering part with the thermal emission part. 
Compared to the model used in paper I, the ZL component at  wavelengths longer than 
3 $\mu$m is slightly brighter. 
In the L band, our adopted models render a $\sim$ 6$\%$ brighter value than
the original model, but still within the uncertainty of the models.
The validity of the adopted ZL spectrum will be compared with observations in section 4. 

\subsection{Integrated star light (ISL)}
The integrated star light (ISL) for stars  fainter than the limiting magnitude composes a portion of the foreground emission. The best
way to estimate the ISL is to sum the brightness of stars in the 2MASS catalog that fall within the beam,
however, the uncertainty of the attitude 
determination and irregular beam pattern elongated along scan path
makes this analysis difficult. While, in paper I, we applied the SKY model \citep{Cohen97}, and assume
a simple $cosec(b)$ law to the model ISL at  three selected fields, in this paper, we obtain the model ISL based on an improved model,
that is, the TRILEGAL Galaxy model \citep{Girardi05}. We further calculate ISL for 12 equally-spaced fields 
along the scan path for the I, J, H, K$_{S}$, L, and L$_{S}$ bands. 

Since the  fluctuation of the ISL is not negligible, we perform 50 Monte-Carlo simulations  
for a 1 square degree field for all TRILEGAL stars fainter than the IRTS limiting magnitude and 
assign their average values to be the TRILEGAL ISL. 
However, the overall uncertainty of the TRILEGAL model is so large, that we must calibrate the  
TRILEGAL ISL by comparing with the 2MASS ISL for the H and K bands. 
The 2MASS ISL is obtained by summing the ISL for the 2MASS stars fainter than the IRTS limiting magnitudes, and that for stars 
fainter than 2MASS limiting magnitudes (15.8 mag for the J band and 15.1 mag for the H band; \citet{Skrutskie06})
which is estimated by adopting the TRILEGAL model. Contribution of the latter part to the 2MASS ISL is only a few percent. 
Fig.1 shows the correlation between the TRILEGAL ISL and the 2MASS ISL for the 12 selected fields. Left and right panels 
indicate the results for the H and K bands, respectively. Horizontal errors in Fig.1 represent the 1 $\sigma$ dispersion of the 
TRILEGAL ISL as a result of these Monte-Carlo simulations and indicate the expected fluctuation of TRILEGAL ISL.
Fig.1 clearly indicates that the 2MASS ISL is brighter than the TRILEGAL ISL 
both for the H and K bands, and the scatter of the 2MASS ISL is consistent with that expected from the
TRILEGAL model. Based on this analysis, we finally obtain the model ISL by multiplying TRILEGAL ISL 
by 1.23 with 8 $\%$ error. 
The ISL for the 1010 fields observed with IRTS was obtained by interpolating the two neighboring model fields 
assuming a  $cosec(b)$ law.  The ISL for 24 of the IRTS bands was estimated by interpolating the model ISL using  
the blackbody spectrum with the same limiting magnitudes as in paper I. 

Besides the model errors, we calculate the model ISL for variation of $\pm$ 0.5 mag in the  limiting 
magnitude and assigned the difference to be peak-to-peak errors. This error is a little lower than the error due to
uncertainty of the model ISL. The model ISL thus obtained renders a more reliable brightness and spatial distribution than that of paper I.

\subsection{Diffuse Galactic} light (DGL)
  The DGL was not taken into account in paper I as a foreground emission, as mentioned in \citet{Mattila06}, 
since no reliable observation of DGL was reported at that time and the contribution of DGL 
to the overall sky brightness was thought to be small.
Recently new observations of the near-infrared DGL have been attained, and we attempt  to estimate the DGL
for the IRTS fields and bands. 
 \citet{Tsumura13c} obtained low resolution spectra of the diffuse sky with AKARI for 
 the wavelengths ranging from 2 $\mu$m  to 5 $\mu$m, 
 detecting a clear correlation between the near-infrared sky brightness and the far-infrared 
 emission (100 $\mu$m, \citet{Schlegel98}).
A PAH band at 3.3 $\mu$m was clearly detected, however, the detection was limited to the Galactic
 plane at $b<15^{\circ}$.
 \citet{Arai15} performed a similar correlation analysis with data from the Low Resolution Spectrometer (LRS) \citep{Tsumura13a}, 
 one of the instruments on the sounding rocket experiment, CIBER (Cosmic Infrared Background ExpeRiment, \citet{Zemcov13}). 
 They also detected a clear correlation with the far-infrared emission 
 for the wavelength range from 0.95 $\mu$m  to 1.65 $\mu$m at high Galactic latitudes. 
 
 Fig.2 summarizes their results. The ratio of the DGL to the far-infrared (100 $\mu$m) emission is shown
 in units of nWm$^{-2}$sr$^{-1}$ / $\mu$Wm$^{-2}$sr$^{-1}$. Filled and open circles represent 
 the result of CIBER and AKARI, respectively.
The  CIBER result indicates scattered star light by interstellar dust, and the AKARI result shows the thermal emission of the fine dust
 particles transiently heated by a single UV photon. We fit the scattered part  with the 
 ZDA04-BC03 \citep{Brandt12} model recommended by \citet{Arai15}, 
 as is shown with the dotted line. As for the thermal part, it is not clear that 
 the AKARI result can be applied to the DGL at high Galactic latitudes, since  \citet{Tsumura13c} reported lower level of thermal
 DGL than expected from Fig.2 at higher Galactic latitudes. Therefore, we first assume that no thermal part exists at IRTS fields  where 
 Galactic latitudes are higher than $40^{\circ}$. The thermal contribution, especially from the 3.3 $\mu$m
 PAH feature will be discussed again in section 4. Adopted values for IRTS bands are shown by open squares
 in Fig.2 for which an error of $\pm 20 \%$ was applied.
  
 We used the far-infrared map compiled by \citet{Schlegel98}, and
 retrieved 100 $\mu$m brightness for a 12 arcmin diameter FOV  for the 1010 IRTS fields. 
Using the ratio of the DGL to the 100 $\mu$m brightness (Fig.2), we are able to obtain DGL for 24 IRTS bands and for 1010
 IRTS fields. 
 
\section{Residual isotropic emission}
We attempt to obtain the residual isotropic emission with same procedure as paper I. First, we subtract the ISL and DGL from 
the observed sky brightness and make a correlation analysis with the model ZL brightness. 
The upper sets of data points in Fig.3 
show typical correlation diagrams at 1.8  $\mu$m both for the Kelsall model (left panel) and the Wright 
model (right panel). For both models, we find an excellent linear correlation for all wavelength bands, and intersection 
of linear fit line at $x=0$ provides the residual  isotropic emission. 
The lower sets of data points in Fig.3 
show the individual residual 
emission for the 1010 IRTS fields at 1.8  $\mu$m after subtracting all foreground
emission components, demonstrating that the residual emission is fairly isotropic.
Since error levels are almost same for two models, it must be noted that there is no clear preference between these two models.

Fig.4 shows the dependence of the residual emission at 1.8  $\mu$m on the Galactic latitudes for the case of the Kelsall model. 
Data points indicate averaged values for 10 degrees along the Galactic longitude. Filled circles and open
circles indicate the result of present work and that of paper I, respectively. 
The large scale structure observed in paper I disappeared, and the data points of the 
present work show random scatter. This improvement is mainly due to our revision of the ISL model.

Fig.5 shows slopes of the linear fit line both for the Kelsall and Wright model. 
In both models, slopes are a few percent larger than 1.0, which is consistent with 
paper I.  We regard this as the deviation of the ZL spectrum from the solar spectrum which 
reflects the physical properties of interplanetary dust. 
The ZL spectrum can be obtained by multiplying the solar spectrum by these values, showing 
reddening of the ZL color at near-infrared wavelengths. 
The observed spectrum smoothly
connects to the ZL spectrum from 0.75 to 1.6 $\mu$m observed with CIBER
and is consistent with the J band brightness of the Kelsall model based on the DIRBE data (see Fig. 9 in
\citet{Tsumura10}).

Fig.6 shows the residual isotropic emission obtained for both the Kelsall and Wright model, with two different sets of error bars.
The inner and outer error bars indicate the random and total error, respectively. 
Aside from the random error, the systematic error makes the spectrum change in the same direction. Random errors include: fitting errors 
from correlation analysis, ISL errors due to limiting magnitudes, calibration errors, and DGL errors. Systematic 
errors are due to model errors of ZL and ISL. As paper I, errors in the ZL model are estimated by interpolating the uncertainties 
in the original model (Table 7 in \citet{Kelsall98}).
ZL model error is the dominant source of error and amounts to $\sim$ 80
$\%$ of total error. 
Table 1 indicates numerical values for the residual emission and their error for two models.

Compared with residual emission in paper I, which is based on the Kelsall model, the peak brightness of the residual emission
of the present work is $\sim$ 11 nWm$^{-2}$sr$^{-1}$ lower. 
The residual emission at  wavelengths longer than
2 $\mu$m is almost the same as that of paper I. 
A flat spectrum for the three shortest wavelength bands 
is a characteristic feature found in this analysis.
The residual emission obtained by adopting the Wright model for the ZL provides $\sim$ 7 nWm$^{-2}$sr$^{-1}$
lower peak value than that obtained by adopting  the Kelsall model but 
the spectral shape is very similar to each other. Fig.6 implies that there exists excess near-infrared isotropic
residual emission independent of the choice of the two ZL models used here. 

Additionally, we make the same analysis including the thermal part of the  DGL. In this case, residual emission at 3.28 $\mu$m PAH feature
is estimated to be $\sim$20 $\%$ ($\sim$ 2.6 nWm$^{-2}$sr$^{-1}$) lower than that of Fig.6, 
causing a sharp absorbing feature in the spectrum of the residual emission. We also find that the sky brightness at 3.28 $\mu$m after
subtracting ZL and ISL shows no correlation with far-infrared sky brightness.
These results favor the non-existence of the PAH 
feature at high Galactic latitudes, however, this is not conclusive given that random noise at 3.28 $\mu$m
is so large. In any case, the contribution of the thermal part of DGL 
is almost negligible and does not significantly change the final result.

Fig.7 shows the breakdown of the emission components whose spectra are obtained as the average brightness
at high ecliptic latitudes ($ \beta>70^{\circ}$) and high Galactic latitudes ($ b>45^{\circ} $) for the case of the
Kelsall model.
The ZL Spectrum indicated is obtained by applying the result of correlation analysis (Fig.5). 
The residual isotropic emission amounts to $\sim$ $1/4$ of
the ZL, and this is comparable with the seasonal variation of the ZL. 

Fig.8 summarizes the observations of the residual isotropic emission for which the Kelsall model is applied
for the ZL. Typical two observations \citep{Cambrecy01, Levenson07} are shown for COBE data.
Solid line indicates the model of the integrated light of galaxies (ILG) by \citet{Totani00}, which is consistent with
deep Galaxy counts \citep{Keenan10}.
The brightness of this work is a little
lower than that in paper I, but still consistent with COBE and AKARI  \citep{Tsumura13d}. 
It must be emphasized that these three satellite observations render consistent
residual emission within error, although the beam size and limiting magnitudes are completely different. 
In the optical region, two new observations are plotted.  \citet{Matsuoka11} re-analyzed the 
Pioneer 10/11 data, where the ZL is negligible near the Jupiter orbit, while 
\citet{Mattila11} attempted to detect the EBL using shadowing effect of dark cloud at high Galactic latitude.
Both results show that the observed background brightness is consistent with the known foreground emission. On the other hand, HST 
observations by \citet{Bernstein07} show fairly bright residual emission at a similar level as this work.

Fig.9 indicates excess brightness over ILG. 
Since some galaxies are already removed in  \citet{Bernstein07} and \citet{Tsumura13d}, 
we subtracted 10  and 50 $\%$ of the ILG (solid line in Fig. 8) from their data in Fig.8, 
respectively, based on ILG magnitude relationship by \citet{Keenan10}.
We find improved consistency between IRTS and AKARI data.
As already mentioned in paper I, the blue Rayleigh-Jeans like spectrum is clearly seen at wavelengths longer than 
1.6 $\mu$m.

\section{Discussion}

It has been suspected that the residual isotropic emission is a part of the ZL since the spectrum
of the residual isotropic emission is similar to that of the ZL \citep{Dwek05}. However, it is not so easy to construct the new
ZL model which includes the residual isotropic emission, since the residual emission is comparable to the seasonal
variation of the ZL. One possible approach is to add the new dust component which has a heliocentric and spherically 
symmetric distribution. It is, however, difficult to maintain 
the spherically symmetric distribution of interplanetary dust of inner solar system due to the perturbation by planets. 
Maintaining the supply of dust 
against Pointing-Robertson drag is also difficult. These considerations imply that the new dust component
must be beyond the orbit of Jupiter. If we attribute the residual isotropic emission to the scattered sunlight by interplanetary dust, 
the brightness extrapolated to visible wavelengths 
amounts to $\sim$ 100 nWm$^{-2}$sr$^{-1}$, assuming a solar spectrum.  
The existence of such a component, however, is not detected with 
Pioneer 10/11 observations \citep{Hanner74}. A recent reanalysis of the Pioneer 10/11 data by \citet{Matsuoka11}
confirms that the residual isotropic emission near the orbit of Jupiter is less than 10 nWm$^{-2}$sr$^{-1}$ 
at 0.44 and 0.64 $\mu$m. These results indicate that the new dust component does not exist, otherwise the new 
component dust has peculiar optical properties.
 
The light of the first stars at the re-ionization epoch is another possible emission source which has been thought
to be an important clue to delineate the star formation history of the universe. However, recent theoretical
works based on high redshift galaxies predict much less contribution of the first stars to the
near-infrared background \citep{Cooray12b, Yue13a}. 

It has been thought that the spatial 
fluctuation of the sky directly provides the characteristic feature of the EBL, since the fluctuation
of the ZL is so low \citep{Pyo12}. Large fluctuations at angles larger than 100 arcsec that 
cannot be explained with the known foreground sources are detected with Spitzer at 3.6 and 4.5 $\mu$m \citep{Kashlinsky05, 
Kashlinsky07a, Kashlinsky07b} and AKARI at 2.4, 3.6 and 4.2 $\mu$m \citep{Matsumoto11}. 
Recent Spitzer \citep{Kashlinsky12, Cooray12a} and AKARI \citep{Seo15} observations
confirm that flat fluctuation spectra extend to degree scales. The spectrum of fluctuation is 
blue Rayleigh-Jeans like \citep{Matsumoto11} and clear spatial correlations are found for both 
AKARI and Spitzer wavelengths. 

\citet{Zemcov14} recently report the result of sounding rocket observations with CIBER. They
detect large fluctuation at 1.1 and 1.6 $\mu$m. In Fig.9 we over-plot the result of  Fig.2 in \citet{Zemcov14}
 as large filled squares. Unit is shown in right ordinate with same dynamic range as excess brightness. 
The spectrum of the fluctuations is similar to the excess brightness, suggesting that the excess 
fluctuation and brightness may be of same origin.  A ratio of the absolute brightness to the
fluctuation,  $I/\Delta I$,  is $\sim$ 30. 
Furthermore, \citet{Zemcov14} find excellent spatial correlation among 1.1, 1.6 and Spitzer 3.6 $\mu$m.
Since no clear signature of redshifted Lyman $\alpha$ is detected, they conclude that the first stars are 
not the source of excess fluctuation. They also exclude direct-collapsed black holes (DCBHs) model \citep{Yue13b}, since
the detected fluctuation is too high and the color is quite different from the prediction.
\citet{Zemcov14} claim that the large portion of the observed fluctuation
can be explained with intra halo light (IHL) and estimate the surface brightness of IHL to be comparable 
to that of ILG. It must be emphasized that the quoted EBL (ILG plus IHL) is still a few times lower 
than that of the residual emission obtained in this work.

At present, there is no definite emission source which can explain the observed excess brightness. 
More observational and theoretical work is needed to delineate the emission source of excess
brightness and fluctuation.

\section{Summary}
The reanalysis of the IRTS data was performed to obtain improved and more 
reliable measurement of the near-infrared residual isotropic emission.
We revised the estimation of the integrated star light due to faint unresolved stars and 
the thermal emission part of the zodiacal light spectrum.
We take the diffuse Galactic light into account based on the recent observations  with AKARI and CIBER.
Besides the Kelsall model for the zodiacal light used in paper I, another model, the Wright model, was examined, too.

The result for the Kelsall model shows peak value of 60 nWm$^{-2}$sr$^{-1}$ 
which is $\sim$ 11 nWm$^{-2}$sr$^{-1}$ lower than that of paper I. This is still considerably brighter than the
integrated light of galaxies (ILG). The result for the Wright model shows peak value of 53 nWm$^{-2}$sr$^{-1}$
which is slightly fainter than that of the Kelsall model. Both models render significant
residual isotropic emission that cannot be explained with known foreground emission sources.

\section{ACKNOWLEDGMENTS}  

The authors thank Shuji Matsuura and Toshiaki Arai for their encouragement and valuable comments. 
Thanks are also due to Edward L. Wright for his courtesy in providing the code of his zodiacal light model.
K.T. was supported by KAKENHI (26800112) from JSPS, Japan.

This publication makes use of data products from the Two Micron All Sky Survey, 
which is a joint project of the University of Massachusetts and the Infrared 
Processing and Analysis Center/California Institute of Technology, funded by 
the National Aeronautics and Space Administration and the National Science Foundation.

\clearpage

\begin{figure}
\epsscale{1.0}
\plotone{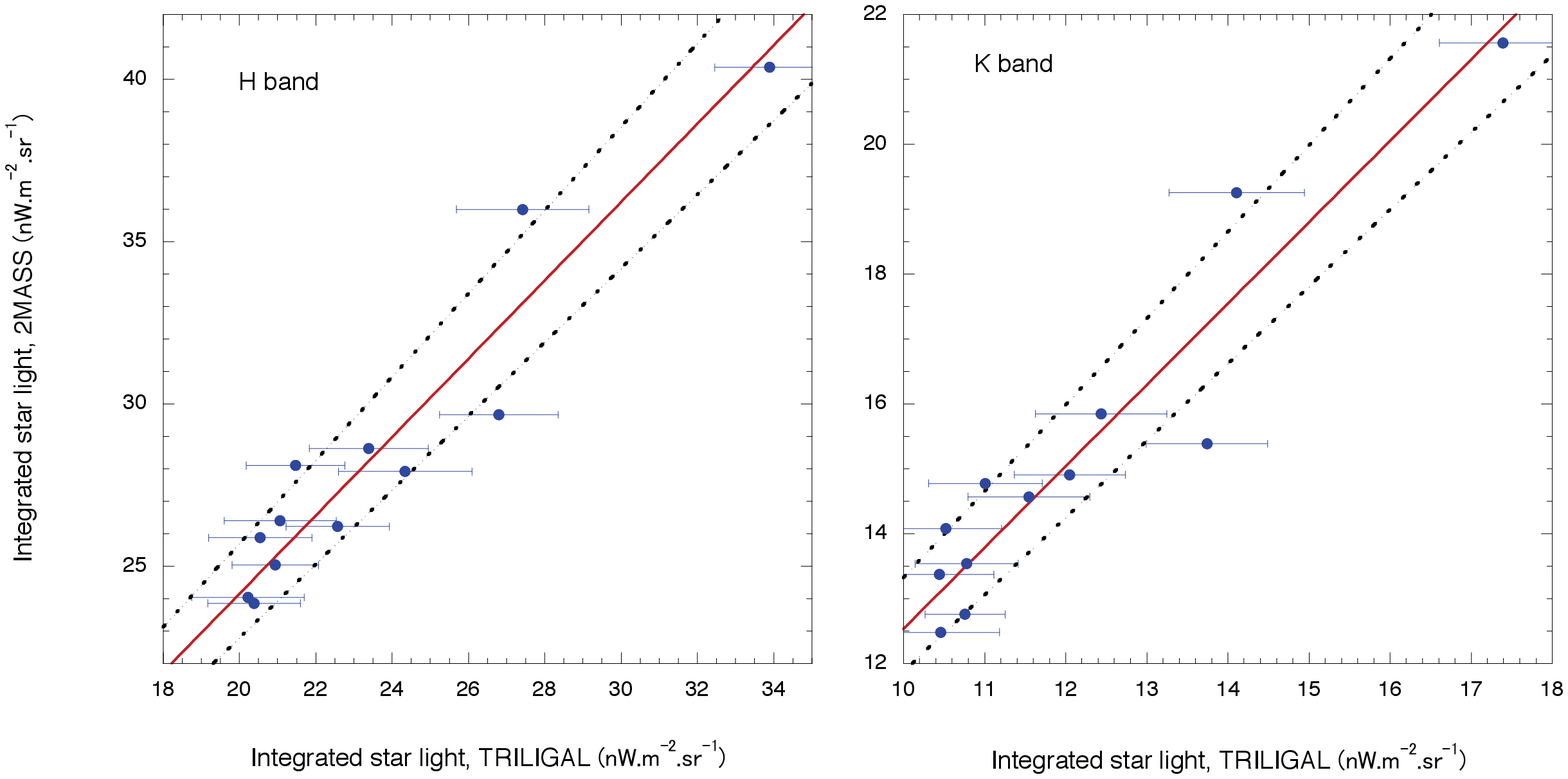}
\caption{
The correlation between the TRILEGAL ISL (the integrated light for stars fainter than the IRTS limiting magnitude based on
the TRILEGAL Galaxy model)
 and  the 2MASS ISL (the integrated light for 2MASS stars fainter than IRTS limiting magnitude and for stars fainter
 than 2MASS limiting magnitude based on the TRILEGAL Galaxy model) for the 12 selected IRTS fields. 
Left and right panels show the case for the H and K band, respectively. Horizontal error bars represent the
standard deviation of the TRILEGAL ISL which was obtained using Monte Carlo simulations.
Straight lines show the best fit for linear correlation, and the dotted lines indicate the adopted
$\pm$1$\sigma$ error.\label{fig1}}
\end{figure}

\clearpage

\begin{figure}
\epsscale{0.8}
\plotone{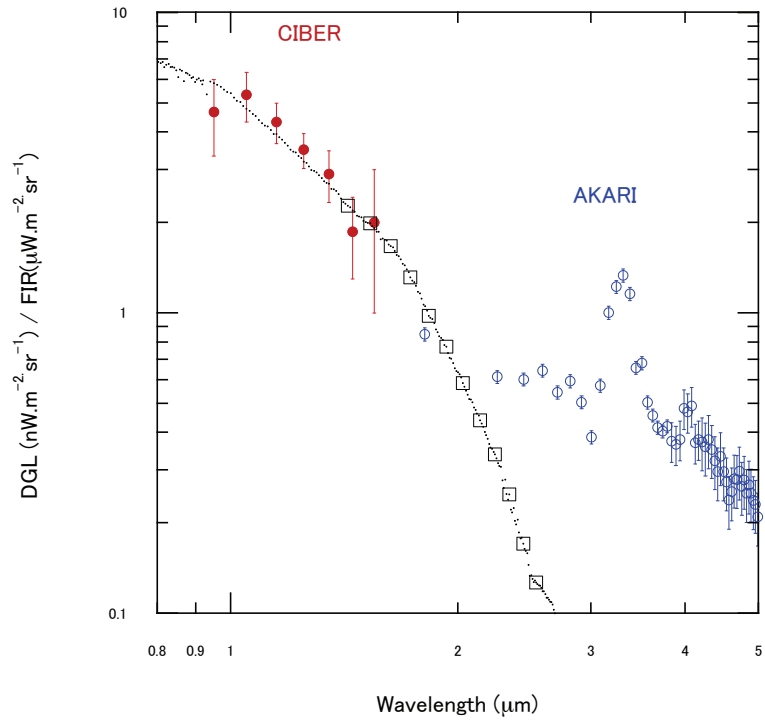}
\caption{The ratio of the DGL to the far-infrared (100 $\mu$m) emission, shown
 in units of nWm$^{-2}$sr$^{-1}$ (DGL)/ $\mu$Wm$^{-2}$sr$^{-1}$ (FIR). Filled and open circles represent 
 the result of CIBER \citep{Arai15} and AKARI \citep{Tsumura13c}, respectively. Dotted line 
 shows the model (ZDA04-BC03 in \citet{Brandt12}) recommended by 
  \citet{Arai15}, and open squares indicate adopted ratio for IRTS bands. \label{fig2}}
\end{figure}

\clearpage

 \begin{figure}
\epsscale{1.0}
\plotone{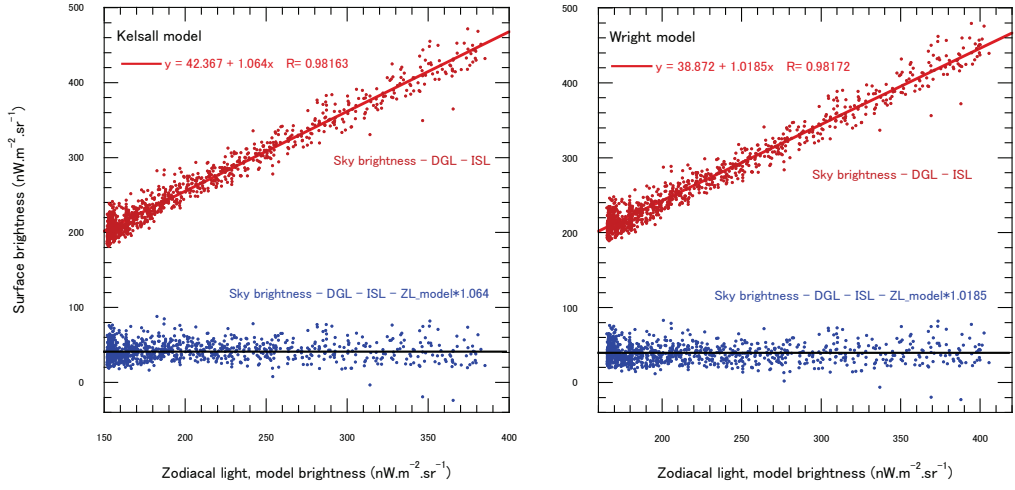}
\caption{
Upper set of data points shows the
correlation diagram between the surface brightness  after subtracting ISL and DGL
from observed sky brightness at 1.8 $\mu$m, and
model ZL brightness. Left and right panels indicate the case for the Kelsall and  Wright 
model, respectively. Solid lines are best fit lines for linear correlation. 
Lower set of data points 
represents individual residual emission after subtracting all foreground emission from observed sky brightness 
in which solid lines show residual emission obtained by linear correlation analysis.  \label{fig3}}
\end{figure}

\clearpage

 \begin{figure}
\epsscale{0.8}
\plotone{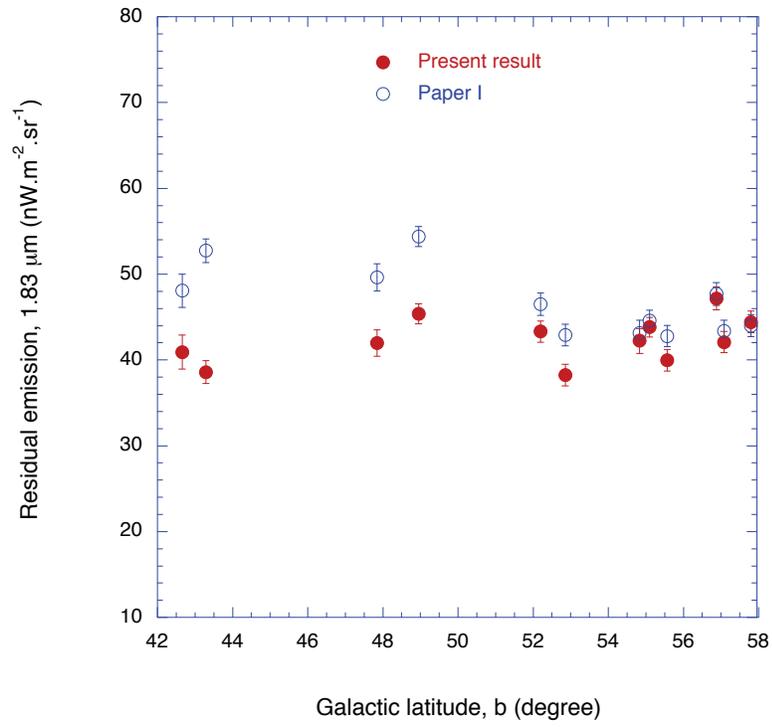}
\caption{
Dependence of residual emission at 1.8 $\mu$m on the Galactic latitudes for the case of the Kelsall model. 
Filled circles and open circles represent the result of this work and that of
paper I, respectively. Data are takes by averaging individual residuals for 10 degrees
along Galactic longitude. 
\label{fig4}}
\end{figure}

\clearpage

 \begin{figure}
\epsscale{0.8}
\plotone{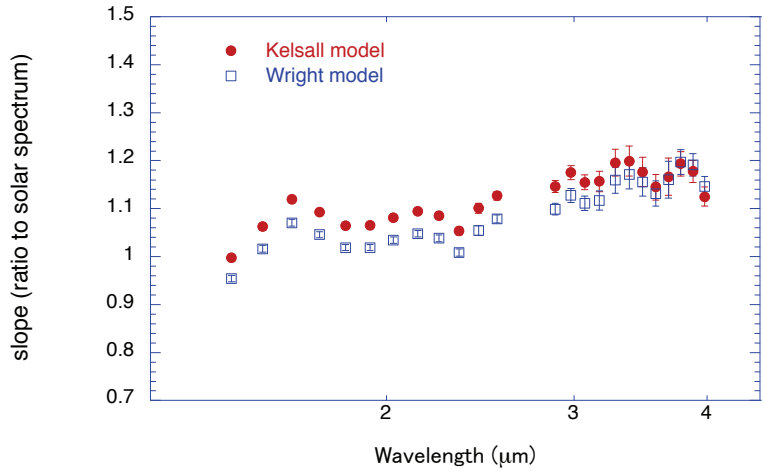}
\caption{
Wavelength dependence of the slopes of the linear fit lines in the upper part of Fig.3. Slopes 
represent ratios to the solar spectrum. Filled and open circles represent the case for the Kelsall model 
and the Wright model, respectively. 
\label{fig5}}
\end{figure}

\clearpage

\begin{figure}
\epsscale{0.8}
\plotone{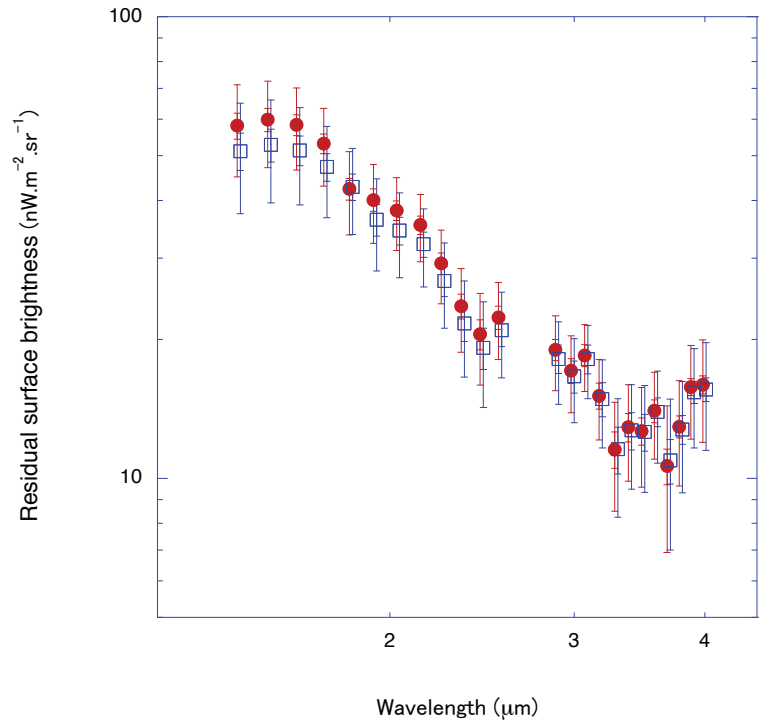}
\caption{Spectra of residual isotropic emission after subtracting ZL, DGL and ISL are shown for the Kelsall (filled circles) 
and the Wright model (open squares), respectively. For clarity, the residual emission of the Wright model is shifted to slightly longer 
wavelength. Two sets of error bars are plotted for each data point. The inner bars represent random errors, 
while the outer bars indicate the total error including systematic errors.\label{fig6}}
\end{figure}

\clearpage

 \begin{figure}
\epsscale{0.8}
\plotone{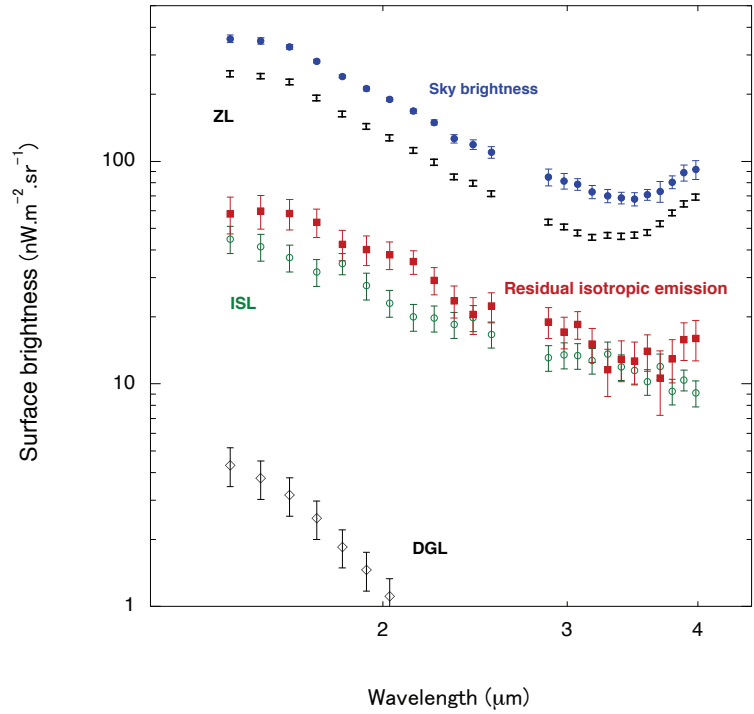}
\caption{A breakdown of the sky brightness at high ecliptic latitudes ($ \beta>70^{\circ}$) 
and high Galactic latitude ($ b>45^{\circ} $)
for the case of the Kelsall model. Filled circles, bars, filled squares, open circles, and open diamonds indicate the observed sky brightness,
zodiacal light (ZL), residual isotropic emission, integrated star light (ISL) and diffuse Galactic light (DGL),
respectively.
 \label{fig7}}
\end{figure}

\clearpage
	
\begin{figure}
\epsscale{1.1}
\plotone{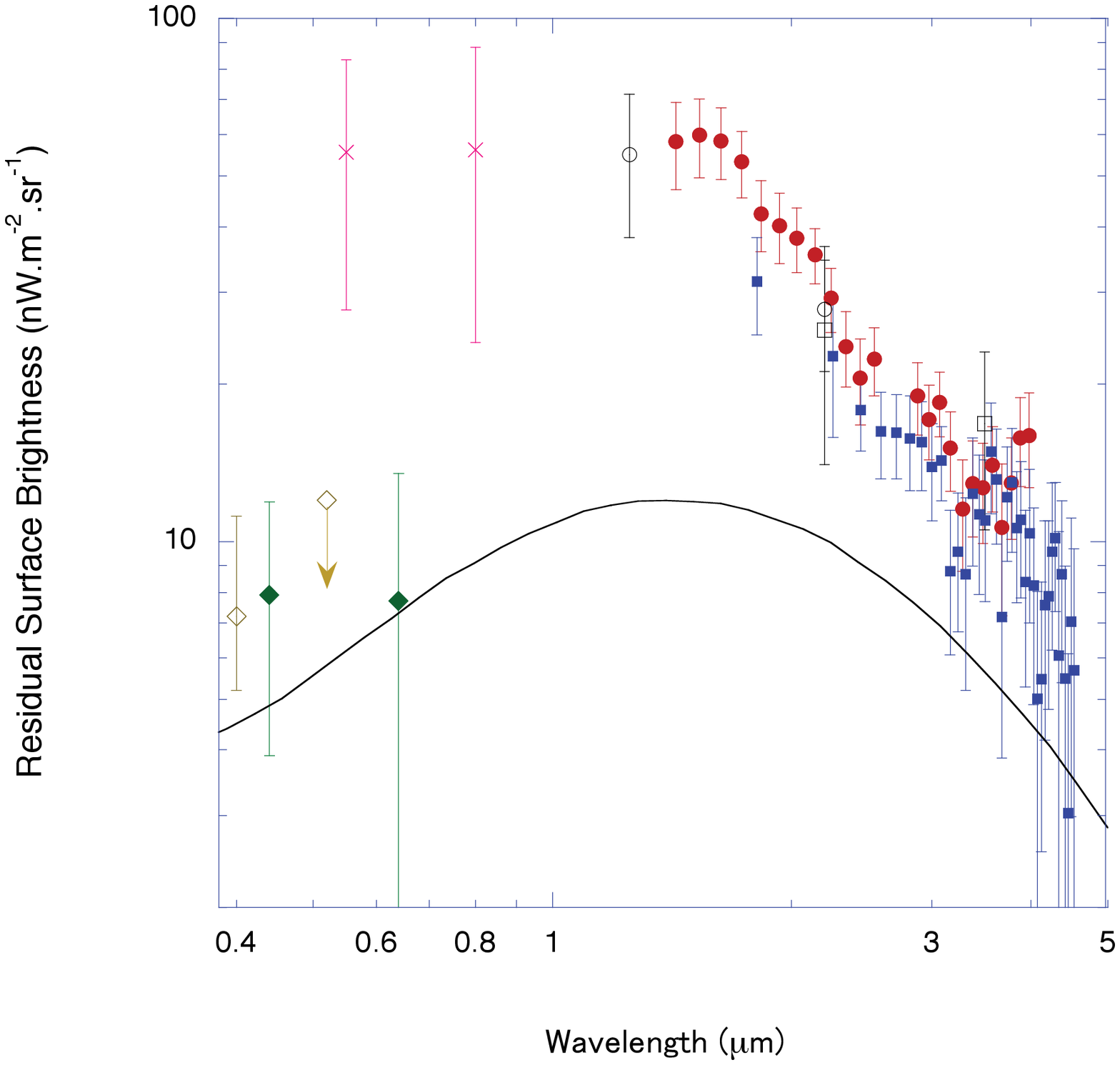}
\caption{
Summary of observations of the residual isotropic emission for the case of the Kelsall model. 
Symbols represent following data: filled circles (IRTS, this work), 
filled squares (AKARI, \citet{Tsumura13d}), open circles (COBE, \citet{Cambrecy01}), open squares (COBE, \citet{Levenson07}),
crosses (HST, \citet{Bernstein07}), filled diamonds (Pioneer 10/11, \citet{Matsuoka11}), and
open diamonds (dark cloud, \citet{Mattila11}). The solid line shows the integrated light of galaxies (ILG) \citep{Totani00}.
\label{fig8}}
\end{figure}

\begin{figure}
\epsscale{1.1}
\plotone{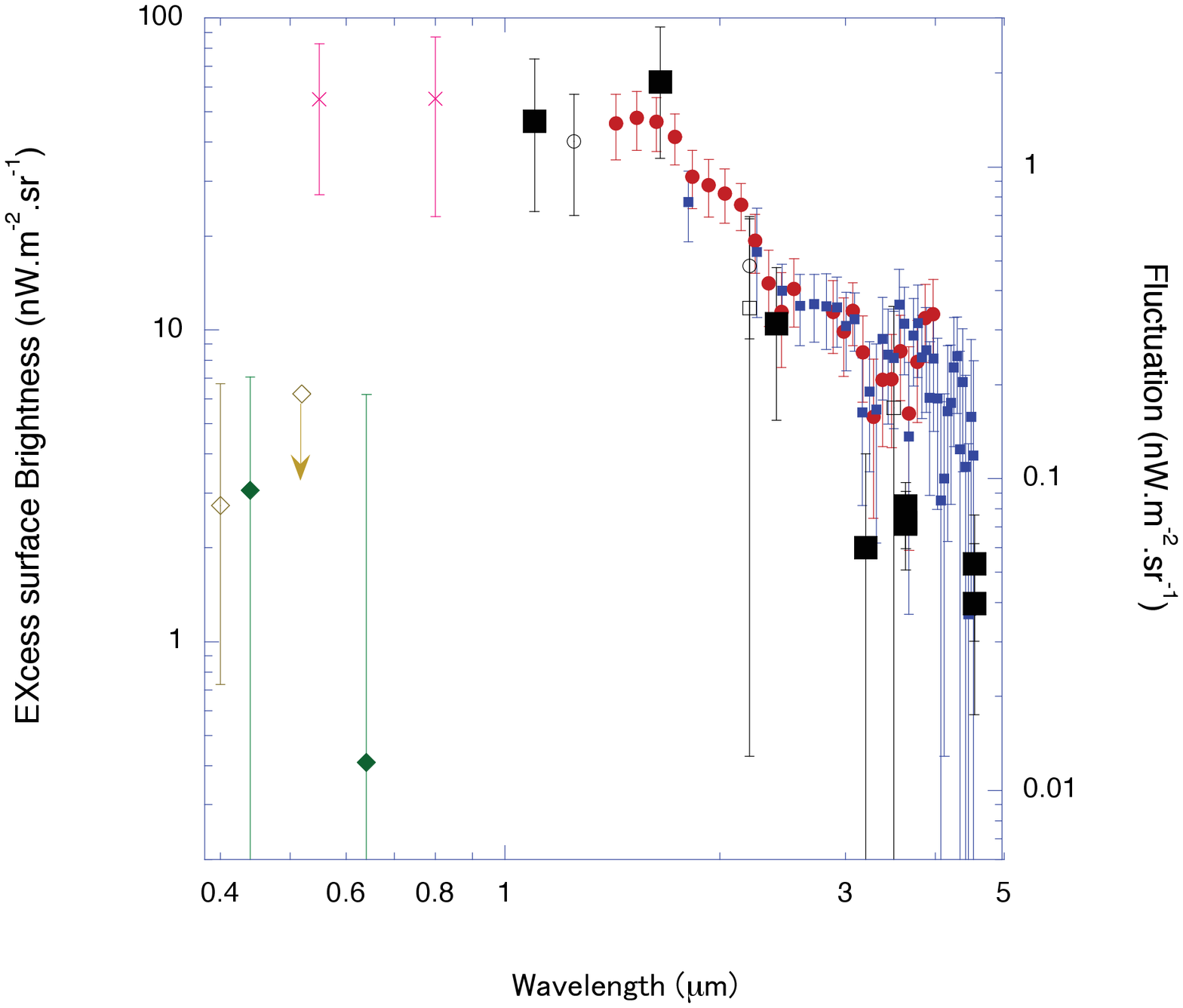}
\caption{
Summary of the excess emission over integrated light of galaxies (ILG). 
Symbols represent the following data: filled circles (IRTS, this work), 
filled squares (AKARI, \citet{Tsumura13d}), open circles (COBE, \citet{Cambrecy01}), open squares (COBE, \citet{Levenson07}),
crosses (HST, \citet{Bernstein07}), filled diamonds (Pioneer 10/11, \citet{Matsuoka11}), and
open diamonds (dark cloud, \citet{Mattila11}). Large filled squares indicate fluctuations of the sky observed 
with Spitzer, AKARI and CIBER \citep{Zemcov14}  with the right ordinate having the 
same dynamic range as the excess brightness. \label{fig9}}
\end{figure}

\clearpage

\begin{table}
\begin{center}
\caption{Surface brightness of residual isotropic emission and its errors
in units of nWm$^{-2}$sr$^{-1}$}
\begin{tabular}{cccccc}
\hline\hline
Wavelength & \multicolumn{2}{c}{Kelsall} & & \multicolumn{2}{c}{Wright} \\
\cline{2-3}\cline{5-6} 
[$\mu$m] & Residual & Total Error & & Residual & Total Error \\
\hline
3.98 & 16.0  & 4.0  & & 15.6  & 4.1  \\
3.88 & 15.8  & 3.6  & & 15.4  & 3.7  \\
3.78 & 12.9  & 3.3  & & 12.8  & 3.4  \\
3.68 & 10.6  & 3.7  & & 10.9  & 3.9  \\
3.58 & 14.0  & 3.0  & & 13.9  & 3.1  \\
3.48 & 12.7  & 3.1  & & 12.6  & 3.3  \\
3.38 & 12.9  & 3.0  & & 12.7  & 3.2  \\
3.28 & 11.6  & 3.1  & & 11.6  & 3.3  \\
3.17 & 15.1  & 3.0  & & 14.9  & 3.2  \\
3.07 & 18.5  & 3.0  & & 18.2  & 3.3  \\
2.98 & 17.1  & 3.2  & & 16.6  & 3.4  \\
2.88 & 19.0  & 3.5  & & 18.2  & 3.7  \\
2.54 & 22.3  & 4.2  & & 20.9  & 4.4  \\
2.44 & 20.5  & 4.6  & & 19.2  & 4.9  \\
2.34 & 23.6  & 4.9  & & 21.7  & 5.1  \\
2.24 & 29.2  & 5.3  & & 26.8  & 5.6  \\
2.14 & 35.4  & 5.9  & & 32.2  & 6.2  \\
2.03 & 38.0  & 6.8  & & 34.4  & 7.2  \\
1.93 & 40.1  & 7.8  & & 36.3  & 8.2  \\
1.83 & 42.4  & 8.7  & & 42.8  & 9.0  \\
1.73 & 53.1  & 10.1  & & 47.3  & 10.6  \\
1.63 & 58.3  & 11.8  & & 51.4  & 12.2  \\
1.53 & 59.9  & 12.7  & & 52.8  & 13.3  \\
1.43 & 58.1  & 13.1  & & 51.2  & 13.7  \\
\hline
\end{tabular}
\end{center}
\end{table}

\clearpage

\end{document}